\documentclass[amsmath,amssymb,graphicx,twocolumn,bibnotes,superscriptaddress]{revtex4-2}

\usepackage{graphicx}
\usepackage{bbm}
\usepackage{physics}
\usepackage{color}
\usepackage{mathrsfs}
\usepackage{amsmath} 
\usepackage{newtxmath}
\usepackage[caption=false]{subfig}
\usepackage[colorlinks=true,citecolor=blue]{hyperref}

\begin{document}
\newcommand{\dif}{\mathrm{d}}

\title{Deterministic and Programmable Fusion for Scalable Generation of Photonic Graph States} 

\author{Yan Li}
\altaffiliation{These two authors contributed equally to this work.}
\affiliation{Center for Quantum Information, Institute for Interdisciplinary Information Sciences, Tsinghua University, Beijing 100084, PR China}

\author{Yunfan Yang}
\altaffiliation{These two authors contributed equally to this work.}
\affiliation{Center for Quantum Information, Institute for Interdisciplinary Information Sciences, Tsinghua University, Beijing 100084, PR China}

\author{Zenghui Bao}
\affiliation{Center for Quantum Information, Institute for Interdisciplinary Information Sciences, Tsinghua University, Beijing 100084, PR China}

\author{Jize Yang}
\affiliation{Center for Quantum Information, Institute for Interdisciplinary Information Sciences, Tsinghua University, Beijing 100084, PR China}

\author{Luyan Sun}
\affiliation{Center for Quantum Information, Institute for Interdisciplinary Information Sciences, Tsinghua University, Beijing 100084, PR China}
\affiliation{Hefei National Laboratory, Hefei 230088, PR China}

\author{Yipu Song}
\affiliation{Center for Quantum Information, Institute for Interdisciplinary Information Sciences, Tsinghua University, Beijing 100084, PR China}
\affiliation{Hefei National Laboratory, Hefei 230088, PR China}

\author{Yukai Wu}
\affiliation{Center for Quantum Information, Institute for Interdisciplinary Information Sciences, Tsinghua University, Beijing 100084, PR China}
\affiliation{Hefei National Laboratory, Hefei 230088, PR China}

\author{Hongyi Zhang}
\email{hyzhang2016@tsinghua.edu.cn}
\affiliation{Center for Quantum Information, Institute for Interdisciplinary Information Sciences, Tsinghua University, Beijing 100084, PR China}
\affiliation{Hefei National Laboratory, Hefei 230088, PR China}

\author{Luming Duan}
\email{lmduan@tsinghua.edu.cn}
\affiliation{Center for Quantum Information, Institute for Interdisciplinary Information Sciences, Tsinghua University, Beijing 100084, PR China}
\affiliation{Hefei National Laboratory, Hefei 230088, PR China}

\begin{abstract}
Entanglement in multiphoton graph states is an important resource with various applications, such as measurement-based quantum computing and memory-less quantum repeaters. While some primitive instances have been successfully prepared, scalable generation of large-scale graph states with programmable connections remains an outstanding challenge. Fusion provides a promising “divide-and-conquer” approach by linking smaller resource states into larger, more complex graphs. However, its practical scalability has been limited by the probabilistic nature of conventional fusion protocols, which necessitates extensive hardware overhead and feed-forward resources to achieve deterministic outcomes. Here, we introduce and experimentally demonstrate a deterministic fusion operation with built-in error mitigation, capable of connecting small time-bin-encoded cluster states—originally generated on-demand from a quantum emitter—into larger, reconfigurable photonic graph states of increased sizes. The fusion operation is realized by nondestructively projecting pairs of flying microwave photonic qubits into Bell states using superconducting circuits with in-situ programmability. Crucially, the protocol features active reset and reuse of matter qubits during graph state generation, thereby circumventing the coherence constraints of the quantum emitter and enabling the generation of genuine multipartite entanglement across 13 photonic qubits. Our results establish a scalable framework for photonic graph-state synthesis, paving the way for fault-tolerant measurement-based quantum computation and distributed quantum networks.

\end{abstract}
\maketitle

Multipartite entangled graph states are essential resources for measurement-based quantum computing, communication, and metrology~\cite{Raussendorf_2001,RevModPhys_pan2012,Lo_2013,Shettell2020,bartolucci_2021,Briegel2009,wang_scalable_2025}. Linear optical techniques utilizing spontaneous parametric down-conversion (SPDC) sources provide a powerful method for generating photonic graph states, enabling several key proof-of-principle experiments~\cite{RevModPhys_pan2012,12_PhotonPRL}. However, due to the inherently probabilistic nature of SPDC, these sources face scalability challenges for large-scale quantum information processing. Atom-like quantum emitters offer a deterministic and efficient alternative for generating single photons and multiphoton entanglement~\cite{Rudolph2009PRL,Besse_2020,Cogan_2023,Thomas_2022,ferreira_2022,Yang_2022,huet_2024,sunada_2024}. Yet, constructing large-scale graph states with the desired topology requires multiple interacting quantum emitters with sufficiently long coherence times~\cite{osullivan_2024,Russo_2019}, a technical challenge that remains difficult to address. To date, genuine multipartite entanglement has been demonstrated in cluster states of up to 12 qubits, yet with limited topological complexity~\cite{Thomas_2022,osullivan_2024}.

Fusion offers a promising strategy for constructing large-scale photonic graph states~\cite{Browne_2005}. By performing entangling measurements on qubits from multiple constant-size entangled resource states, fusion enables the synthesis of larger, topologically complex many-body entangled states~\cite{Briegel2009,bartolucci_2021,Hilaire_2023}. Recently, fusion of resource states from quantum emitters has been demonstrated using two-photon interference and destructive projective measurements in linear optical setups~\cite{Thomas_2024,Lodahl2025}. However, the projective measurements are inherently probabilistic~\cite{Browne_2005}, with a maximum success probability of 50\%. Additionally, photon loss in the optical path or during detection further diminishes the fidelity of the operation~\cite{PhysRevX.5.041007}. To address the non-deterministic nature of linear optics-based fusion, ancilla qubits and repeated operations are necessary~\cite{Loock2014,Hilaire_2023}, which increases the resource overhead for quantum information processing. Furthermore, a successful two-photon interference requires perfect overlap of the temporal, spectral, and spatial modes of the input photons. This imposes a stringent limitation on the flexibility of the fusion operation and, consequently, on the resulting entanglement scale and structure.


\begin{figure*}[!htbp]
\centering
\includegraphics[width=1\linewidth]{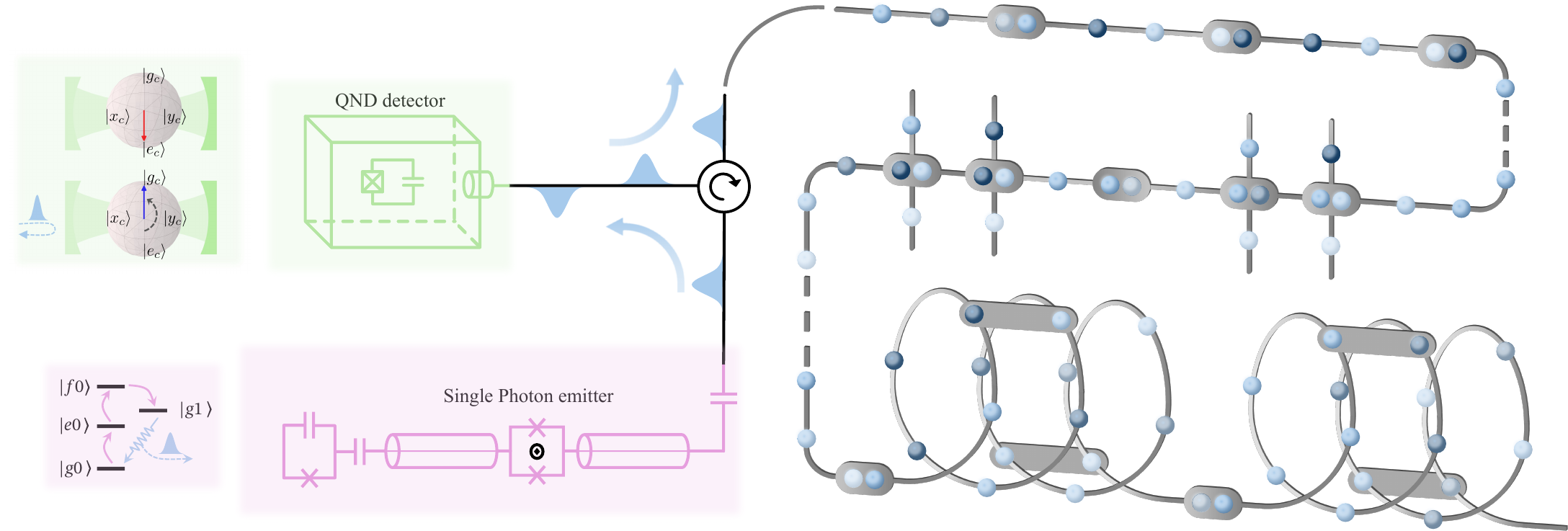}
\caption{\textbf{Scalable and reconfigurable generation of microwave photonic graph states via deterministic fusion.} A single-photon emitter (pink) generates sequential photonic flying qubits to form small cluster states, routed to a quantum non-demolition (QND) parity detector (green) for deterministic fusion (grey bars). In situ frequency tuning of the photonic qubits via a tunable resonator enables programmable fusion between arbitrary qubit pairs, allowing the synthesis of complex graph topologies. The matter qubit in the emitter can be reset between the generation of individual small cluster states, thereby localizing errors within subgraph units (blue vertices, fading gradient). Likewise, the qubit in the QND detector can be reset between successive fusion operations to mitigate error propagation arising from its finite coherence time. This architecture enables scalable, high-fidelity, and reconfigurable generation of large-scale, high-dimensional photonic graph states. Schematic illustrations of the operating principles of the single-photon emitter and the QND detector are shown to their respective left.}
\label{setup}
\end{figure*}

To address these challenges, we propose a deterministic and programmable fusion operation based on circuit/cavity quantum electrodynamics (cQED) setup~\cite{Duan_2004,Blais_2021,Reiserer_2015}. Leveraging cavity-assisted interactions between flying photonic qubits, cQED setup has been successfully used to realize photon-photon gates~\cite{Hacker_2016, Niemietz_2021, Reuer_2022}, generate entanglement between remote qubits~\cite{Reiserer_2014, Neuzner_2016, Kurpiers_2018}, and perform quantum non-demolition (QND) detection of flying photonic states~\cite{Reiserer_2013, Kono_2018, Besse_2018}. Notably, by performing QND parity measurement~\cite{Besse_2020_parity,Wang_2022}, two flying photons can be deterministically prepared in Bell states, potentially enabling a deterministic and QND fusion operation~\cite{Browne_2005}.

In this work, we demonstrate the first deterministic fusion operation for flying microwave photonic qubits using superconducting quantum circuits~\cite{Blais_2021}. A single-photon emitter, which deterministically generates itinerant time-bin photonic qubits or multiphoton entangled cluster states, is coupled to a photon detector that performs a QND parity measurement on the incoming photons.
We first show the deterministic fusion of two initially uncorrelated time-bin qubits through selective parity measurements. Then, we generate photonic cluster states and fuse them into larger, higher-dimensional states in a programmable manner. Additionally, we present repetitive fusion operations by resetting matter qubits during graph state generation, allowing the creation of a genuine multipartite entangled 11-vertex cluster state encoded with 13 photonic qubits.
The fusion operation demonstrated here exhibits scalability and feasibility, setting the stage for large-scale graph state-based quantum information processing.

\bigskip

\textbf{Fusion operation with circuit QED setup.}

The experimental setup is illustrated in Fig.~\ref{setup}. A single-photon emitter~\cite{LI_2023} and a QND single-photon detector~\cite{Kono_2018,Wang_2022}, both consisting of a transmon qubit dispersively coupled to a microwave resonator, are cascade-connected via a microwave circulator. The single-photon emitter generates itinerant microwave photonic qubits or one-dimensional (1D) cluster states by sequentially transferring population from the second excited state to microwave photons via a resonator-assisted $f0g1$ transition~\cite{Kurpiers_2019}. The frequency of the emitted photonic qubits can be programmed on demand by adjusting the magnetic flux through a superconducting quantum interference device (SQUID) embedded at the center of the resonator~\cite{Besse_2020,LI_2023}. The photonic states are analyzed with a homodyne setup and reconstructed with a maximum likelihood method~\cite{Eichler2012}. Details about photonic state generation and reconstruction can be found in Sections III and IV of the Supplementary Information.

The proposed fusion operation relies on QND parity measurement of flying photons. As illustrated in the left panel of Fig.~\ref{setup}, the QND photon detector comprises a superconducting microwave cavity that is dispersively coupled to a transmon qubit~\cite{Kono_2018,Wang_2022}. When the transmon qubit is prepared in a superposition state $\ket{+_c} =1/\sqrt{2}(\ket{g_c}+\ket{e_c})$, where $\ket{g_c}$ ($\ket{e_c}$) represents the ground (excited) state, a flying photon reflected from the cavity imprints a $\pi$ phase shift on the qubit, resulting in an entangled state of the form $\ket{+_c}\ket{0}+\ket{-_c}\ket{1}$, where $\ket{0}$ and $\ket{1}$ represent the vacuum and Fock-1 states, respectively~\cite{Duan_2004, Kono_2018}. If another photon is sent to the cavity, the system evolves into the entangled state  $\ket{+_c}(\ket{00}+\ket{11})+\ket{-_c}(\ket{01}+\ket{10})$. At this stage, a projective measurement on the transmon qubit will collapse the two photons into a Bell state. 
As illustrated in Fig.~\ref{setup}, if the two photons are initially in different graph states, the fused Bell state serves as a redundantly encoded vertex that inherits all the graph bonds of the two photons~\cite{Browne_2005}, thereby merging the two graph states into a new one. It is important to emphasize that the measurement outcome of the transmon only affects the encoding basis of the fused photonic qubits, without altering the connections of the graph states. This ensures that the fusion operation is deterministic without any heralding.

\begin{figure*}[!htbp]
\centering
\includegraphics[width=0.9\linewidth]{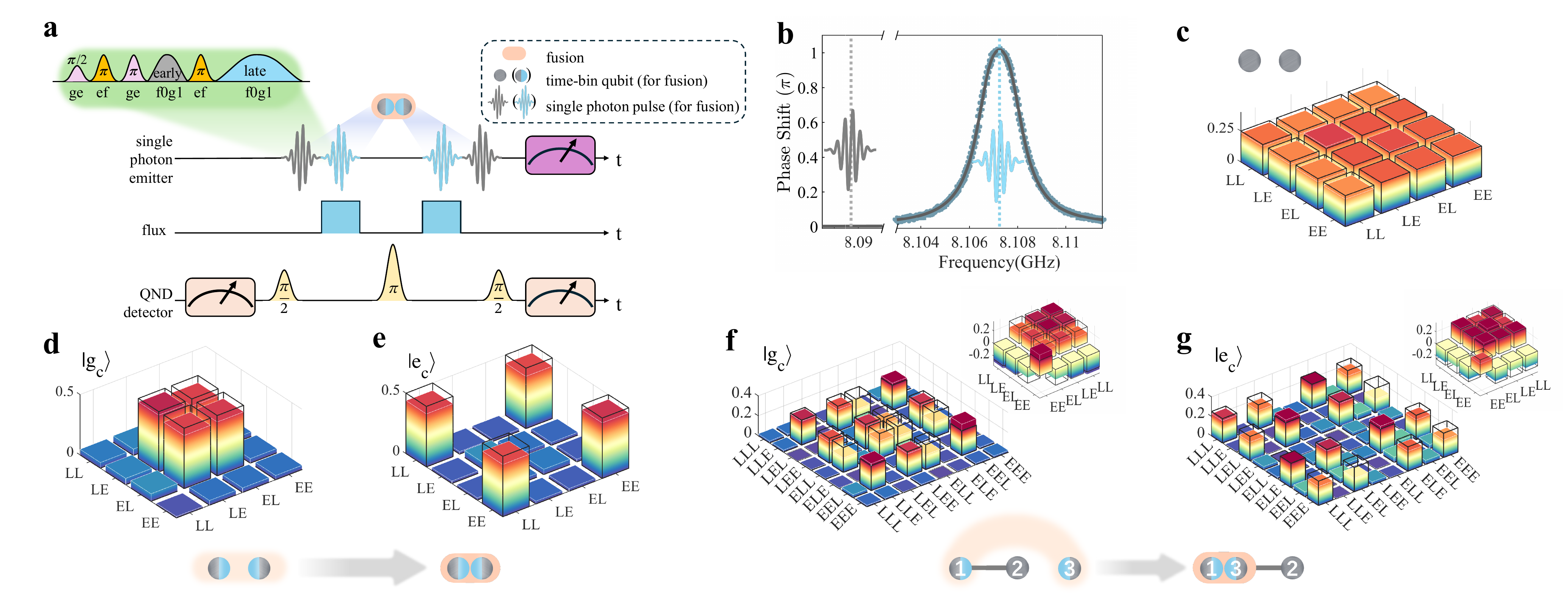}
\caption{\textbf{Deterministic fusion operation.} \textbf{a}, Pulse sequence demonstrating the deterministic fusion of two time-bin encoded photonic qubits. A single-photon emitter sequentially generates two time-bin encoded photonic qubits. The fusion operation is performed via a controlled-phase flip between the QND detector and one of the modes of each photonic qubit. Details of the photonic qubit generation and fusion process are provided in the Methods section. The frequency of the single-photon pulses is tunable in situ by controlling the magnetic flux across the SQUID. The photonic states are analyzed using a homodyne setup for quantum state tomography. \textbf{b}, Conditional phase flip as a function of the input single-photon frequency, which is close to $\pi$ at resonance. The frequencies used in the experiment, on- and off-resonance, are marked by dashed lines. \textbf{c}, The reconstructed density matrix of the uncorrelated input photonic qubits. \textbf{d} and \textbf{e}, The reconstructed density matrix of the two-qubit photonic state after the fusion operation, conditioned on the transmon qubit being projected onto the states $\ket{g_c}$ and $\ket{e_c}$, respectively. \textbf{f} and \textbf{g}, Fusing qubit 1 in a cluster state with isolated qubit 3 generates a redundantly encoded two-vertex cluster state. The corresponding three-qubit density matrix is reconstructed, conditioned on the transmon being projected onto $\ket{g_c}$ and $\ket{e_c}$, respectively. The insets in the upper-right corner display the reduced density matrices of the redundantly encoded two-qubit state, obtained by projecting the density matrices onto the corresponding subspaces. The density matrices of the corresponding ideal states are shown with hollow caps.}
\label{fusion}
\end{figure*}

Programmable fusion can be realized by leveraging the frequency-tunable single photons and the finite bandwidth of the QND detector. As shown in Fig.~\ref{fusion}\textbf{b}, input photons with a proper frequency can be entangled with the transmon qubit through the conditional $\pi$ phase shift. In contrast, if the frequency of the incoming photons falls outside the QND detector's bandwidth, they are simply reflected from the cavity without a conditional phase shift, thereby preventing any entanglement with the transmon. By dynamically tuning the frequency of the photonic modes, we can selectively fuse specific photonic modes within a train of photonic pulses, thus realizing programmable fusion operation, as detailed in the Methods.

\textbf{Deterministic fusion.}
  
To demonstrate the deterministic fusion operation, we generate two photonic qubits and send them to the QND detector. The photonic qubit is defined with time-bin encoding, where a single photon exists in a superposition of two temporal modes~\cite{Kurpiers_2019, ilves_demand_2020,LI_2023}. 
A photon in the early or late temporal mode is denoted as $\ket{E} \equiv \ket{10}$ or $\ket{L} \equiv \ket{01}$, respectively. 
Within the single-photon subspace, two time-bin qubits can be fused through parity measurements on one of the temporal modes in each qubit, which is realized by in situ tuning the frequency of the photon modes (Methods). The pulse sequence is shown in Fig.~\ref{fusion}\textbf{a}. Conditioned on the transmon qubits projected in $\ket{e_c}$ or $\ket{g_c}$, the two photonic qubits are prepared in Bell states $1/\sqrt{2}(\ket{EE}+\ket{LL})$ or $1/\sqrt{2}(\ket{EL}+\ket{LE})$. 
Fig.~\ref{fusion}\textbf{c}, \textbf{d}, and \textbf{e} show the reconstructed density matrix $\rho_{\text{exp}}$ for the photonic qubits without and with the fusion operation. One can see that the initially uncorrelated qubits are prepared into Bell states by the fusion operation. We have the Bell state fidelity $F= \Tr(\rho_{\text{exp}}\rho_{\text{ideal}})$ of $89.94\%$ and $88.16\%$ conditioned on the transmon qubits in $\ket{g_c}$ or $\ket{e_c}$, respectively.
We estimate that the fusion operation introduces an infidelity of approximately 7.8\% to 9\%, which could be further reduced through improvements in qubit coherence times and optimization of the QND detector bandwidth. Additional details are provided in Section VI of the Supplementary Information.

\bigskip

Notably, the fusion errors can be effectively mitigated by leveraging the QND nature of parity measurements for the photonic qubits. Since the fusion operation preserves the parity subspace, repeated measurements enable statistical enhancement of fidelity by suppressing uncorrelated errors~\cite{Zurek_2009}. For instance, if the transmon is measured in $\ket{e_c}$, the photonic qubits are projected into the subspace spanned by \{$\ket{EE}, \ket{LL}$\}. 
A subsequent measurement yielding a state in the orthogonal \{$\ket{EL}, \ket{LE}$\} subspace indicates an error occurred during the fusion. Therefore, a sequence of cascaded parity measurements on the same vertex offers a straightforward and effective strategy for suppressing the fusion error probability.

When the fused photonic qubits are part of graph states, all bonds attached to them can be preserved through the redundant encoded vertex in the new graph. The fused vertex is redundantly encoded in the basis spanned by \{$\ket{EE}, \ket{LL}$\} or \{$\ket{EL}, \ket{LE}$\} depending on whether the transmon qubit is measured in $\ket{e_c}$ or $\ket{g_c}$, respectively. 
To illustrate, we sequentially generate two photonic qubits in a cluster state and an additional isolated photonic qubit. We selectively fuse one of the photonic qubits from the cluster state with the isolated qubit. As shown in the upper panel of Fig.~\ref{fusion}\textbf{f} and \textbf{g}, the density matrix of the resulting three-qubit state is reconstructed, conditioned on the measurement outcome of the transmon qubit. Fidelities of 82.53\% conditioned on $\ket{g_c}$ and 84.72\% conditioned on $\ket{e_c}$ are achieved, compared with the corresponding ideal states.

To verify the redundant encoding, we extract the logical qubit state by projecting the three-qubit density matrices onto the corresponding encoding subspace, as shown in the insets of Fig.~\ref{fusion}\textbf{f} and \textbf{g}. The resulting reduced density matrices exhibit excellent agreement with the expected two-qubit cluster states—conditioned on the transmon qubit being measured in both $\ket{g_c}$ and $\ket{e_c}$—with the fidelities of 88.23\% and 87.86\%, respectively. These results also confirm the deterministic nature of the fusion operation.

\begin{figure}[!htbp]
\centering
\includegraphics[width=0.99\linewidth]{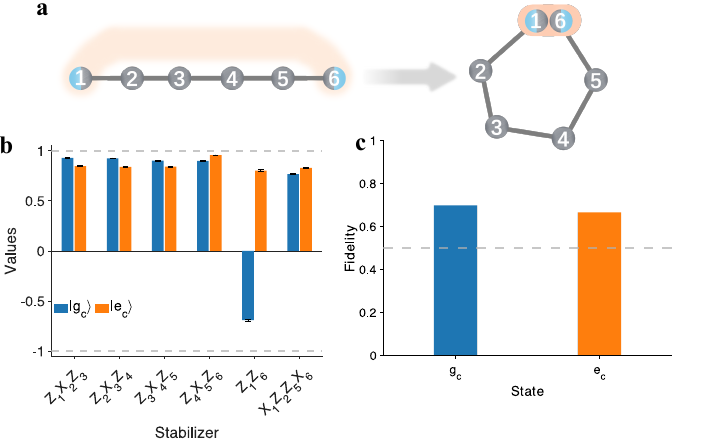}
\caption{\textbf{Programmable graph state generation through selective fusion operations.} \textbf{a}, By utilizing programmable photon generation and fusion operations, a ring graph state with five vertices can be generated. This is achieved by first creating a linear cluster state with 6 qubits, followed by fusing qubit 1 and qubit 6 to form a new vertex. \textbf{b} shows the measured stabilizers. The error bars represent one standard deviation. The dashed gray lines indicate the ideal stabilizer values. \textbf{c}, Fidelity of the experimentally prepared ring graph with respect to the ideal state, computed from the reconstructed process maps. The dashed gray line marks the fidelity threshold of 0.5 for genuine multipartite entanglement.}
\label{program}
\end{figure}

\bigskip
\begin{figure*}[!htbp]
\centering
\includegraphics[width=0.85\linewidth]{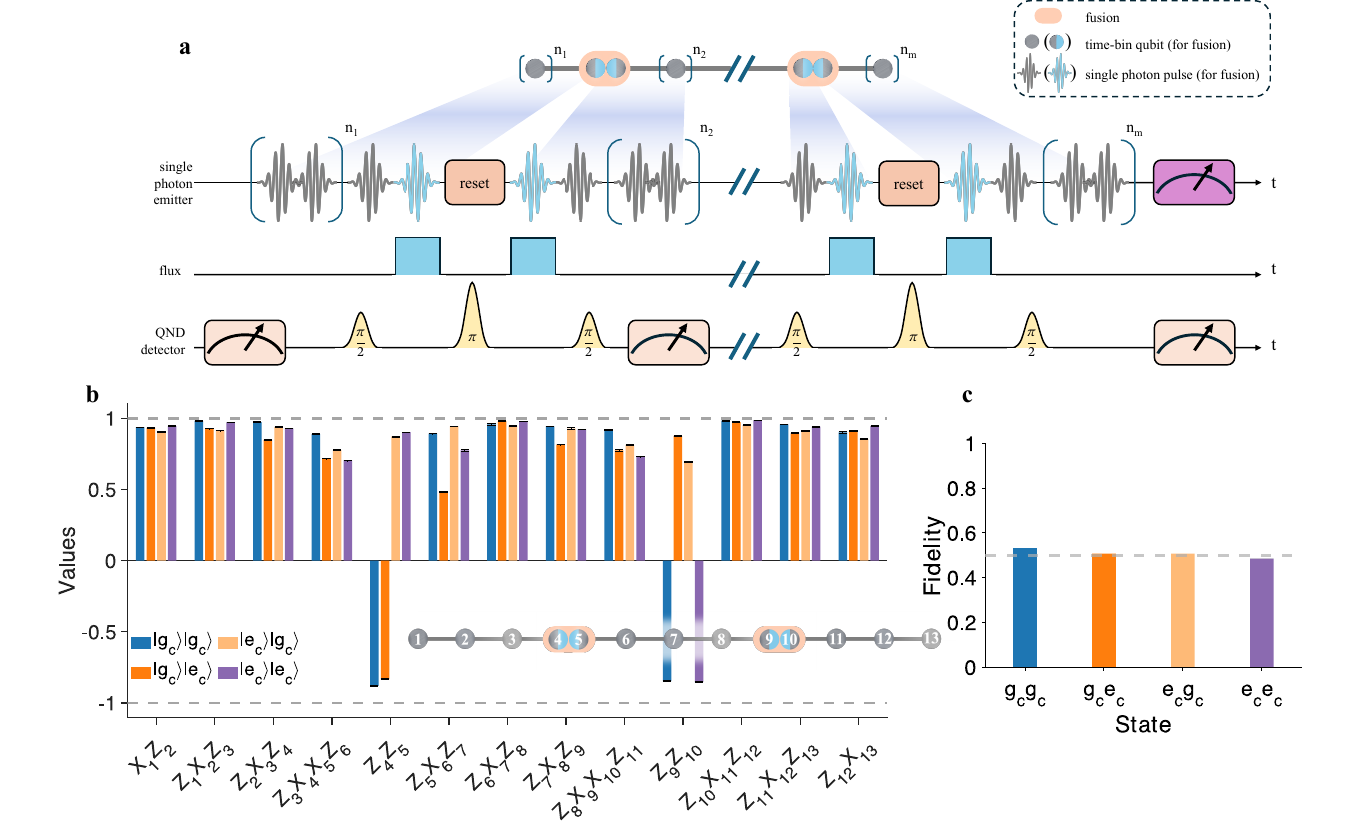}
\caption{\textbf{Scalable graph state generation through repetitive fusion.} \textbf{a}, Pulse sequence for the sequential generation of multiple linear cluster states and the repetitive fusion operation. The shadowed pulse segments represent the preparation of linear cluster states for time-bin qubits, as described in the Methods section. To localize potential errors, the transmon qubits in both the single-photon emitter and the QND detector are reset between the preparation of adjacent cluster states and between consecutive fusion operations, respectively. \textbf{b}, Measured stabilizers for a linear graph state containing 11 vertices, redundantly encoded with 13 photonic qubits, as shown in the inset. This state is generated through two-time fusion of three linear cluster states, comprising 4, 5, and 4 time-bin qubits, respectively. The error bars represent one standard deviation. The dashed gray lines indicate the ideal stabilizer values. \textbf{c}, Fidelity of the repetitively fused linear cluster state with respect to the corresponding ideal state, computed from the reconstructed process maps. The dashed gray line marks the fidelity threshold of 0.5 for genuine multipartite entanglement.}
\label{scalable}
\end{figure*}

\textbf{Programmable fusion.}

The fusion operation can be applied programmatically between any pair of qubits within the resource states. In this context, we demonstrate the creation of a 2D ring graph, a structure that cannot be generated using only a single-photon emitter, by selectively fusing vertices in a 1D cluster state. As shown in Fig.~\ref{program}\textbf{a}, we generate a linear cluster state with six qubits with the single-photon emitter and fuse the first and last qubits to form the ring graph with five vertices.

Due to the significant computational overhead in reconstructing the full density matrix for graph states containing more than four qubits, we use two methods to evaluate the experimental states. First, we characterize the experimental states by performing quantum process tomography on the elementary operations that constitute the generation protocol—namely, photon emission and fusion. Using the reconstructed process maps, we simulate the full state preparation and reconstruct the resulting graph states via the matrix product operator (MPO) formalism. This approach provides a powerful framework for quantitatively comparing the experimentally prepared state with its ideal counterpart through the reconstructed density matrix $\rho_{\text{exp}}$~\cite{Besse_2020,osullivan_2024}. Full details of the process tomography and the MPO-based state reconstruction procedure are provided in Sections VII and VIII of the Supplementary Information. 
In Fig.~\ref{program}\textbf{c}, we report the fidelities of the reconstructed fused ring graph states, conditioned on the control qubit being in $\ket{g_c}$ and $\ket{e_c}$, respectively. The fidelities with respect to the ideal cluster state $\rho_{\text{ideal}}$ are $F = Tr(\rho_{\text{exp}}\rho_{\text{ideal}})$ = 0.70 for $\ket{g_c}$ and 0.67 for $\ket{e_c}$.

Considering that the process maps may overestimate the fidelity of the prepared states~\cite{sunada_2024}, we further measure the stabilizers (Methods) of the prepared states to compare them with the ideal graph states. 
The stabilizers of a graph state are defined as $S_{i} = X_{i} \prod_{i \in V_{i}}Z_{j}$, where $V_{i}$ represents the neighbors of vertex $i$, and $X_{i}$ and $Z{_i}$ denote the Pauli operators. For our redundantly encoded graph, the stabilizer can be translated to a regular graph, up to a local unitary transformation on the redundant photonic qubits~\cite{Hilaire_2023}. The results are shown in Fig.~\ref{program}\textbf{b}. It demonstrates a large overlap between the experimental state and the corresponding ideal state, a circle graph containing five vertices. To detect the entanglement in the experimental state, we use an entanglement witness based on the measured stabilizers~\cite{Toth_2005}, $\mathcal{W}_{G_n} \equiv (N-1)\mathbbm{1} - \sum_{k}S_k^{G_n}$, where $N$ represents the number of qubits, $S_k^{G_n}$ are elements of the stabilizer group for graph $G_n$. Genuine multipartite entanglement can be certified whenever $\mathcal{W}_{G_n} < 0$. 
From the stabilizer measurement shown in Fig.~\ref{program}\textbf{b}, we calculate $\mathcal{W}_{G_6}^{g_c}= -0.107\pm0.017$, $\mathcal{W}_{G_6}^{e_c}=-0.114\pm0.012$, demonstrating genuine entanglement for the deterministically fused graph state.

We note that the programmable fusion operation offers significant flexibility for the reconfigurable generation of graph states. This flexibility is further enhanced when combined with the reconfigurable cluster state generation of a single-photon emitter~\cite{Thomas_2022, huet_2024}. In the Extended Data Fig.~\ref{caterpillar_state}, we showcase the generation of a five-vertex caterpillar state. Those results highlight that the fusion operation is powerful in generating photonic graph states with increasing dimensionality and topological complexity.

\bigskip

\textbf{Scalable graph state generation through repetitive fusion.}

When using quantum emitters to generate graph states, part of the matter qubit errors 
would accumulate in the sequentially generated photonic qubit and propagate across the entire graph state~\cite{Rudolph2009PRL,osullivan_2024,Thomas_2024,Besse_2020}. This leads to a rapid loss of fidelity in the multiphoton entanglement, and thus imposes an inherent limitation on the size of the graph state entanglement that can be generated~\cite{sunada_2024}. To address this challenge, we utilize quantum emitters to generate multiple small-scale resource states, which can be consecutively merged into large graph states through repetitive fusion operations, as illustrated in Fig.~\ref{scalable}\textbf{a}. The quantum emitter can be fully reset between generating adjacent resource states, thereby localizing matter qubit errors to a small segment of the graph and preventing their accumulation, effectively mitigating the fidelity constraints imposed by the finite coherence times of the quantum emitter.
 In Section VIII of the Supplementary Information, we present a theoretical model that demonstrates how the repetitive fusion protocol mitigates error accumulation and enhances the scalability of graph-state generation. Our analysis reveals an optimal segment length that balances fusion-induced errors against accumulated emitter errors, thereby maximizing the overall state fidelity. For our current experimental parameters, this optimum occurs at a segment length of about 5 qubits.

We demonstrate a repetitive fusion on three linear resource states with $n_1=n_3=4$ and $n_2=5$, generating a linear cluster state containing 11 vertices encoded with 13 photonic qubits. Figure~\ref{scalable}\textbf{b} shows the measured stabilizers of the fused graph state, conditioned on the two separate measurement results of the transmon qubit in the QND detector. As with the case of a single fusion operation, the projected transmon states determine the encoding subspace for each fused vertex but do not affect the topology of the generated graph state. A detailed discussion can be found in the Methods section. We characterize the experimental states using both quantum process tomography and stabilizer-based measurements. Fig.~\ref{scalable}\textbf{c} shows the fidelities of the fused graph states with respect to their corresponding ideal states, computed from the reconstructed process maps. The fidelities are 0.53, 0.51, 0.51, and 0.49 for the four post-selected cases conditioned on the qubit in $\ket{g_c}\ket{g_c}$, $\ket{e_c}\ket{g_c}$, $\ket{g_c}\ket{e_c}$, and $\ket{e_c}\ket{e_c}$, respectively.

Regarding stabilizer measurements, we note that at vertices where fusion occurs, the measured stabilizer values exhibit varying deviations from the ideal expectations. These discrepancies can be attributed to the differing readout errors for the $\ket{g_c}$ and $\ket{e_c}$ qubit states, as well as imperfections in the conditional $\pi$-phase shift implemented during QND detection. Despite these discrepancies, the majority of measured stabilizers exhibit significant overlap with the corresponding ideal state, indicating that two-time fusion effectively connects the three cluster states into a larger graph with the desired topology. 
Furthermore, since the matter qubit is decoupled from the photonic states after each cluster state is generated, we do not observe a noticeable reduction in stabilizers from the initial to the final vertex of the graph, indicating that the matter qubit error can be effectively localized.

As an additional verification, we detect genuine multipartite entanglement through stabilizer measurements. Prominently, we certify genuine multipartite entanglement in the fused 13-qubit state conditioned on the transmon qubit being projected to the ground state twice, with $\mathcal{W}_{G_{13}}^{g_c,g_c}= -0.059\pm0.009 < 0$. The certification of genuine multipartite entanglement in the other measurement branches did not surpass the required threshold, primarily due to the limited readout fidelity of the $\ket{e_c}$ state. We note that the imperfect fusion constitutes the primary limiting factor in scaling multipartite entanglement, a challenge that can be mitigated through optimization of the QND detector or by implementing the error-mitigating fusion protocol. In addition, improved entanglement witnesses or full quantum state tomography~\cite{sunada_2024,osullivan_2024} would enhance the certification of multipartite entanglement at larger scales.

\bigskip

\textbf{Conclusion and outlook.}

To conclude,  we have demonstrated a deterministic and programmable fusion operation for flying microwave photonic qubits, enabling reconfigurable expansion of small-scale cluster states from a single quantum emitter into larger entangled resources~\cite{Lobl2025transforminggraph}. This approach offers scalable generation of graph states through deterministic photonic qubit generation and fusion operation, which—unlike probabilistic linear-optical methods—ensures high hardware efficiency. By iteratively fusing emitter-generated subgraphs, matter-qubit error propagation is effectively mitigated, enhancing the fidelity and scalability of multiphoton entanglement. Furthermore, the system’s programmability allows in situ tuning of graph state topologies, paving the way for implementing error correction protocols~\cite{Benjamin_23}.

In the current experiment, we have realized a deterministic fusion operation, but not deterministic graph-state generation. The latter would require a quantum emitter with near-unity photon extraction efficiency. Even short of unity, a high single-photon generation rate is essential to achieve a sufficiently high multiphoton coincidence rate, which is necessary to observe large-scale multipartite correlations within a practical measurement time~\cite{Thomas_2022}. In our present setup, the average single-photon rate incident on the QND detector is approximately 0.35, which currently limits our ability to efficiently probe large photonic graph states. We emphasize, however, that this limitation is not intrinsic to the protocol. The single-photon rate can be substantially improved through enhancements in source design—such as increasing emission efficiency, minimizing optical losses, and optimizing temporal pulse shaping—as well as by refining the overall experimental architecture. Further discussion of these potential improvements is provided in Section IV of the Supplementary Information.

A straightforward extension of the current work involves cascading multiple QND detectors. On the one hand, as previously discussed, such a setup would enable repeated parity measurements, thereby supporting error-mitigation strategies and enhancing fusion fidelity. On the other hand, a key limitation of relying on a single QND detector is that it restricts simultaneous fusion operations to non-crossing pairs of qubits—that is, the fused vertex pairs must form a non-overlapping, non-interleaving matching consistent with the linear ordering of the cluster. This topological constraint prevents the parallel creation of arbitrarily long-range or entangled connections, limiting the expressivity of the generated graph state. Employing additional QND detectors relaxes this constraint and enables the generation of richer graph connectivity beyond the schematic illustrated in Fig.~\ref{setup}.

Moreover, a key feature of the fusion operation described here is its intrinsic QND nature, in contrast to conventional destructive fusion schemes that fix the graph state topology upon qubit measurement. This QND character enables dynamic rewiring of graph states, which permits the fused vertices to be reconfigured on demand, either to meet specific algorithmic requirements or to implement customized quantum error correction protocols. Beyond these advantages, the QND capability supports the development of hybrid quantum interconnects. For example, photonic qubits can be entangled via fusion while preserving their coherent coupling to matter qubits, opening avenues for quantum memory, quantum transduction, and distributed quantum processing. Importantly, while our implementation operates in the microwave domain, the underlying principles are broadly applicable to optical platforms~\cite{Nowak2021nc,Niemietz_2021}. When combined with frequency-tunable single-photon sources~\cite{PhysRevLett.131.170801,Yang_2024}, systems such as single-atom cavity QED~\cite{Reiserer_2015,Reiserer_2013} could adopt these protocols to realize a robust and scalable platform for photonic quantum information processing, with direct applications in measurement-based quantum computing and quantum networks.

\bibliography{ref_fusion24.bib}

\bigskip

\section*{methods}

\textbf{Generation of time-bin encoded photonic qubits.} 
With the single-photon emitter, microwave single photons can be prepared by transferring the second excited state $\ket{f}$ population to microwave photons through a resonator-assisted Raman transition (the $f0g1$ transition)~\cite{Kurpiers_2019,LI_2023}, after which the photons escape from the resonator to the transmission line as propagating single photons. 

To prepare a time-bin encoded photonic qubit, we define the quantum state within the qubit \{$\ket{g},\ket{e}$\} subspace. Then we can sequentially convert the $\ket{e}$ and $\ket{g}$ populations to single photons in early and late time bins, respectively, mediated by $\ket{f}$ state. The pulse sequence is shown in the upper panel of Fig.~\ref{fusion}\textbf{a}. Starting from an equal superposition of $\ket{g}$ and $\ket{e}$, we first transfer the $\ket{e}$ state population to $\ket{f}$ and generate the first time-bin
$$
\ket{g}+\ket{e} \xrightarrow{\pi_{\text{ef}}} \ket{g}+\ket{f} \xrightarrow{\pi_{\text{ge}}} \ket{e}+\ket{f} \xrightarrow{\pi_{\text{f0g1}}} \ket{e}\ket{0}+\ket{g}\ket{1}.
$$
Then we generate the second time-bin
$$
\ket{e}\ket{0}+\ket{g}\ket{1} \xrightarrow{\pi_{\text{ef}}}  \ket{f}\ket{0}+\ket{g}\ket{1} \xrightarrow{\pi_{\text{f0g1}}}  \ket{g}(\ket{10}+\ket{01}).
$$
At this stage, the quantum state encoded in the transmon is converted into a time-bin photonic qubit. By in situ tuning the resonator frequency, the frequency of the emitted photons in either time bin can be selectively brought into or out of resonance with the QND detector. This capability enables a programmable fusion operation.

\textbf{Generation of linear cluster states.} 
A linear cluster state can be generated from a single-photon emitter~\cite{Rudolph2009PRL,sunada_2024}. The underlying principle involves performing repetitive photon emissions conditioned on the transmon state of the single-photon emitter, thereby forming one-dimensional chains of photonic qubits in a cluster state.

In this work, we modify the protocol slightly to prepare time-bin encoded photonic qubits in linear cluster states. The process begins with a quantum non-demolition (QND) measurement and post-selection to initialize the transmon qubit into the ground state. The pulse sequence for generating one-dimensional cluster states from time-bin qubits is illustrated in the Extended Data Fig.~\ref{cluster}.
In general, the pulse sequence enables conditional generation of a single photon in the early time bin ($\ket{E}$) if the transmon is in the excited state ($\ket{e}$), or in the late time bin ($\ket{L}$) if the transmon is in the ground state ($\ket{g}$). For instance, after the first round of conditional photon emission, the system is in the superposition state $\ket{g}\ket{L} + \ket{e}\ket{E}$. Over $N-1$ rounds of conditional photon emission, the transmon qubit and the $N-1$ time-bin encoded photonic qubits are prepared into a one-dimensional cluster state. Finally, the transmon qubit is disentangled from the photonic qubits by converting the transmon qubit into the $N_{\text{th}}$ photonic qubit at the end of the sequence, resulting in an $N$-qubit photonic cluster state.

\textbf{Fusion of time-bin encoded qubits.} 
In the main text, we discussed the protocol for deterministic fusion operation for single-photon encoded qubits. It is essentially the same for the fusion of time-bin encoded photonic qubits, only if the QND parity detector detects one of the temporal modes for two time-bin qubits. We consider two sequentially generated time-bin photonic qubits sent to the QND detector. As illustrated in Fig.~\ref{fusion}\textbf{a}, we control the later (earlier) photonic mode of the first (second) qubit on-resonance with the QND detector, while other modes are kept off-resonance. An additional $\pi$ rotation is applied to the transmon qubit in the QND detector between the incoming of the two resonant single-photon pulses as a spin-echo to refocus the transmon qubit coherence.
If the transmon is measured in $\ket{e_c}$, it indicates an odd photon number in the two resonant modes. Correspondingly, we would have the reflected photonic two-qubit states in a Bell state $\ket{EE}+\ket{LL} \equiv \ket{1010} + \ket{0101}$. On the other hand, if the transmon is measured in $\ket{g_c}$, we would have $\ket{EL}+\ket{LE}\equiv \ket{1001} + \ket{0110}$ for the reflected photonic qubits. As such, we prepare the initially uncorrelated time-bin qubits to Bell states conditioned on the projection results of the transmon in the QND detector.

Similar to the single photon case, even though we have different Bell states conditioned on the projective measurement result of the transmon qubit, this post-determined two-photon entangled state can be used as a redundant encoding as a vertex of the graph states~\cite{Hilaire_2023,Thomas_2024}, which is known as the fusion operation. The fused qubit inherits all the cluster state bonds of the two qubits that were input.

\textbf{Repetitive fusion.}

The repetitive fusion of sequentially generated subgraphs relies on reusing the transmon qubit of the QND parity detector. If the transmon is reset to the ground state, $\ket{g_c}$, after each fusion operation, the QND detector becomes ready for the next fusion event, enabling continuous fusion with a single QND detector. Alternatively, this process can be streamlined by utilizing the QND measurement of the transmon qubit state. By measuring the transmon qubit state before each fusion operation, we can infer its state based on the measurement outcome from the previous round, thereby eliminating the need for an explicit resetting procedure. Consequently, the encoding subspace of the fused photonic qubit pair is determined by the projective measurement outcomes of the transmon qubit, both from the current and previous rounds.

More specifically, for the pulse sequence shown in Fig.~\ref{scalable}\textbf{a}, if the transmon qubit in the QND detector is in the ground state, $\ket{g_c}$, before the fusion operation, the fused photonic state will be encoded in the {$\ket{EE}, \ket{LL}$} ({$\ket{EL}, \ket{LE}$}) subspace, depending on whether the transmon is projected into $\ket{e_c}$ ($\ket{g_c}$). Conversely, if the transmon is initially in the excited state, $\ket{e_c}$, before the fusion operation, the fused photonic state will be encoded in the {$\ket{EL}, \ket{LE}$} ({$\ket{EE}, \ket{LL}$}) subspace, depending on whether the transmon is projected into $\ket{e_c}$ ($\ket{g_c}$). This scenario is equivalent to initializing the transmon in the QND detector in the ground state, $\ket{g_c}$, before each fusion operation, but with a different encoding subspace. Therefore, by knowing the qubit's state before the fusion operation, it becomes unnecessary to reset the transmon qubit explicitly.
For the two-time fusion scenario illustrated in Fig.~\ref{scalable}, four possible outcomes arise based on the results of two consecutive measurements on the transmon: $\ket{e_c}\ket{e_c}$, $\ket{e_c}\ket{g_c}$, $\ket{g_c}\ket{e_c}$, and $\ket{g_c}\ket{g_c}$. The corresponding encoding subspaces for the two fused vertices are $\{\ket{EE},\ket{LL}\}\otimes\{\ket{EL},\ket{LE}\}$, $\{\ket{EE},\ket{LL}\}\otimes\{\ket{EE},\ket{LL}\}$, $\{\ket{EL},\ket{LE}\}\otimes\{\ket{EE},\ket{LL}\}$, and $\{\ket{EL},\ket{LE}\}\otimes\{\ket{EL},\ket{LE}\}$. These outcomes also determine the expectation values of the stabilizer, as depicted in Fig.~\ref{scalable}\textbf{b}.

\textbf{Stabilizer measurements.} 
For graph states containing more than 4 photonic qubits (i.e., 8 photonic temporal modes), reconstructing the full density matrix with QST is technically challenging due to the high computational resource cost. Instead, we use the stabilizers to characterize the entanglement structure of the graph state. 

The stabilizers of a given graph state are defined as $S_{i} = X_{i} \prod_{i \in V_{i}}Z_{j}$, where $V_{i}$ represents the neighbors of vertex $i$. For the graph sections that compose a redundantly encoded vertex, it is equivalent to the regular graph up to a local unitary transformation on the redundant photonic qubits~\cite{Hilaire_2023}. To extract the expectation values for a given stabilizer, we reconstruct the reduced density matrix of the corresponding subset of qubits by measuring the quadratures of their photonic modes. Details can be found in Section III of the Supplementary Information. The expectation value of the corresponding operators can be calculated from the reduced density matrix.
 Error bars for the stabilizer expectation values are calculated from the parametric bootstrap resampling~\cite{Jonathan_2009}.

\textbf{Stabilizer-based entanglement witness.} 
By measuring the expectation values of the stabilizers, we can construct an entanglement witness for the graph states, which allows for the detection of entanglement without requiring full state tomography. We employ the N-partite genuine entanglement witness~\cite{Toth_2005}, defined as $\mathcal{W}_{G_n} \equiv (N-1)\mathbbm{1} - \sum_{k}S_k^{G_n}$, where $N$ represents the number of qubits, $S_k^{G_n}$ are elements of the stabilizer group corresponding to the graph $G_n$. For the ideal state, we would have $\mathcal{W}_{G_n} = -1$.

To validate the applicability of this witness for our redundantly encoded graph state, we evaluate the maximal squared overlap between the state and the set of biseparable states. Notably, the redundantly encoded graph state is local-Clifford equivalent to a standard graph state~\cite{Hilaire_2023}. As shown in Ref.~\cite{Toth_2005}, for such states, a negative expectation value $\langle \mathcal{W}_{G_n} \rangle<0$ is sufficient to confirm the presence of genuine N-partite entanglement.

\bigskip

\textbf{Acknowledgments:} This work was supported by the Quantum Science and Technology-National Science and Technology Major Project (2021ZD0301704), the National Natural Science Foundation of China under Grant Nos. 12374472 and 92165209, the Tsinghua University Initiative Scientific Research Program, and the Ministry of Education of China.

\bigskip

\textbf{Author Contributions:} H.Y.Z. conceived the idea and proposed the experiment. Y.L. and Y.F.Y. collected and analyzed the data with H.Y.Z. Y.L. and Z.H.B. prepared the sample. H.Y.Z. and Y.L. wrote the manuscript. All authors contributed to the discussions and production of the manuscript. L.M.D. and H.Y.Z. supervised the project.

\bigskip

\textbf{Competing interests:} The authors declare that there are no competing interests.

\bigskip
\textbf{Data Availability:} The data that support the findings of this study are available from the authors upon request.

\setcounter{figure}{0}
\renewcommand{\figurename}{Extended Data Fig.}
\renewcommand{\tablename}{Extended Data Table}

\begin{figure*}[!htbp]
    \centering
    \includegraphics[width=0.6\linewidth]{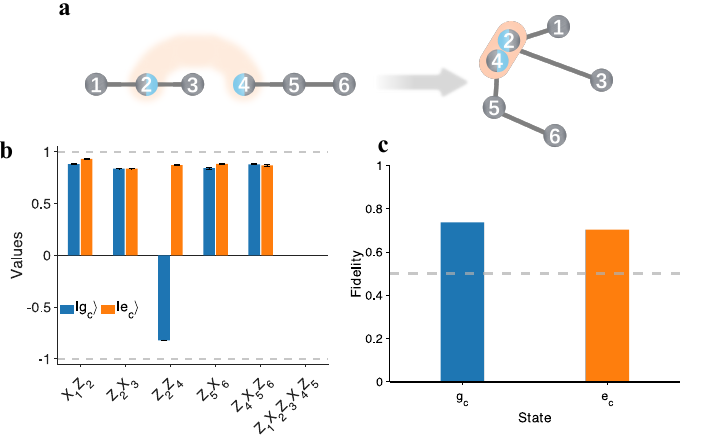}
    \caption{\textbf{Generation of a five-vertex caterpillar state.} \textbf{a}, A caterpillar graph state with five vertices is generated by programmable single-photon generation and fusion operations. This is achieved by first creating two linear cluster states with 3 qubits, followed by fusing qubit 2 and qubit 4 to form a new vertex. \textbf{b} shows the measured stabilizers. Note that the stabilizer $Z_1X_2Z_3X_4Z_5$ is not measured, as quantum state tomography of the five-qubit density matrix would be prohibitively costly due to the large dimension of the Hilbert space. The error bars represent one standard deviation. The dashed gray lines indicate the ideal stabilizer values. \textbf{c}, Fidelity of the experimentally prepared caterpillar state with respect to the ideal state, computed from the reconstructed process maps. The dashed gray line marks the fidelity threshold of 0.5 for genuine multipartite entanglement.}
\label{caterpillar_state}
\end{figure*}

\begin{figure*}[!htbp]
    \centering
    \includegraphics[width=0.7\linewidth]{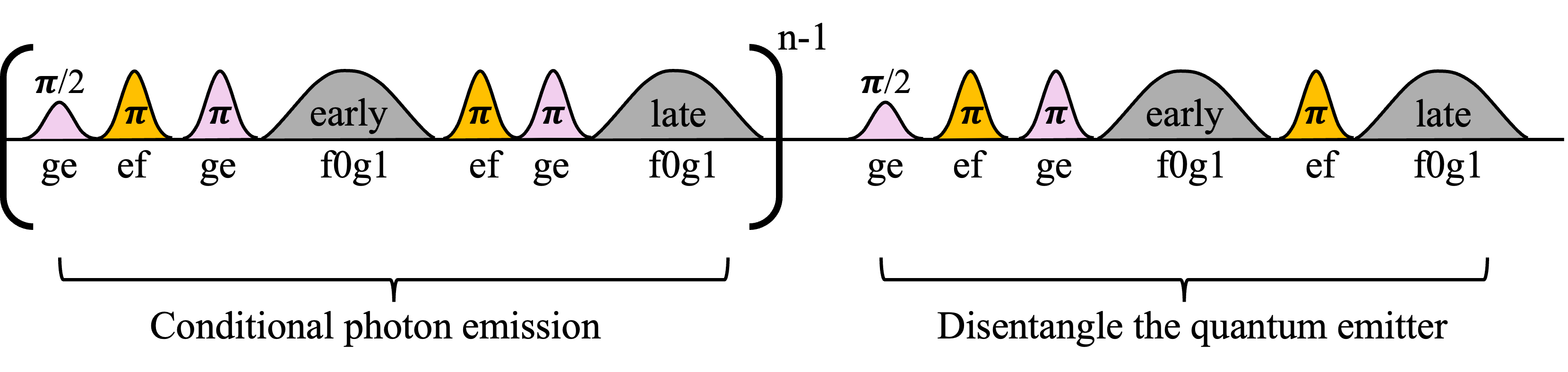}
    \caption{\textbf{Pulse sequence for the generation of 1D cluster state with the single-photon emitter.} }
\label{cluster}
\end{figure*}

\clearpage

\end{document}